\documentclass[12pt,dvips,final]{article}

\usepackage{times}		
\usepackage{graphicx}		
\usepackage{color}		
\usepackage{fancyheadings_b}	
\usepackage{wrapfig}		
\usepackage{bm}			
\usepackage{natbib}		
\usepackage[english]{babel}

\newcommand{\etal}	{\mbox{et al.\ }}%
\newcommand{\eg}	{\mbox{e.g.}}%
\newcommand{\tsim}	{\ensuremath{\sim}}%
\newcommand{\Lya}	{\ensuremath{\mbox{\textrm{Ly}}\alpha}}%
\newcounter{ion}
\newcommand{\ion}[2]	{\setcounter{ion}{#2}\mbox{#1\,{\small\@\Roman{ion}}}}%
\newcommand{\farcs}	{\mbox{\ensuremath{.\mkern-5mu^{\prime\prime}}}}%
\newcommand{\lesssim}	{\mbox{\rlap{\hbox{\lower3pt\hbox{\ensuremath{\sim}}}}\raise2pt\hbox{\ensuremath{<}}}}%
\newcommand{\gtrsim}	{\mbox{\rlap{\hbox{\lower3pt\hbox{\ensuremath{\sim}}}}\raise2pt\hbox{\ensuremath{>}}}}%

\newcommand{\JWST}      {\emph{JWST}}%
\newcommand{\HST}       {\emph{HST}}%
\newcommand{\SDSS}      {\emph{SDSS}}%

\definecolor{navy}{rgb}{0,0,0.63}
\definecolor{lblue}{rgb}{0.2,0.8,1}
\definecolor{lightcyan}{rgb}{0.5,1,1}
\definecolor{dgreen}{rgb}{0.0,0.5,0.0}
\definecolor{lgreen}{rgb}{0.56,0.93,0.49}
\definecolor{gold}{rgb}{0.90,0.75,0.00}
\definecolor{orange}{rgb}{1.0,0.5,0.0}
\definecolor{maroon}{rgb}{0.6,0,0.2}
\definecolor{grey}{rgb}{0.1,0.1,0.1}

\newlength{\txw}
\newlength{\txh}

\begin{document}
\addtocounter{page}{-2}


\pagestyle{empty}
\setlength{\textwidth}{8.5in}
\setlength{\textheight}{11.0in}
\setlength{\oddsidemargin}{-1.0in}
\setlength{\evensidemargin}{\oddsidemargin}
\setlength{\parindent}{0pt}

\noindent\hspace*{0.0in}\begin{minipage}[t]{8.5in}
   \vspace*{-1.35in}\includegraphics[width=8.5in]{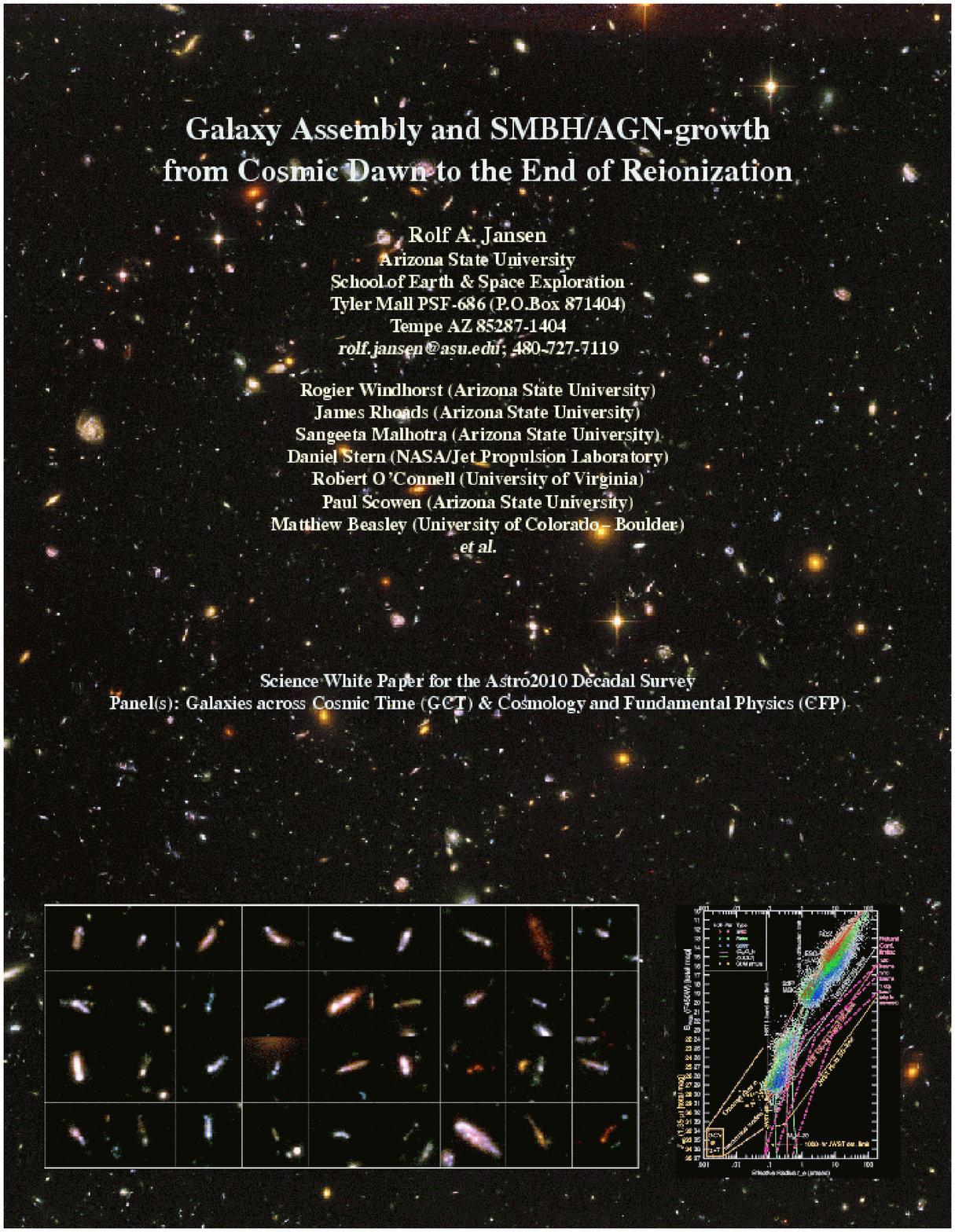}
\end{minipage}
\newpage


\setlength{\textwidth}{6.5in}
\setlength{\textheight}{9.0in}
\setlength{\oddsidemargin}{0pt}
\setlength{\evensidemargin}{\oddsidemargin}
\setlength{\topmargin}{-0.45in}
\setlength{\headheight}{0.55in}
\setlength{\headsep}{0.10in}
\setlength{\topskip}{12pt}
\setlength{\footskip}{14pt}
\setlength{\parindent}{1em}

\setlength{\headrulewidth}{1pt}
\lhead{\large\slshape Galaxy Assembly and SMBH/AGN-growth\\[-6pt]
{\footnotesize\upshape Science White Paper for the Astro2010 Decadal Survey}\\[-14pt]}
\chead{}
\rhead{\large\upshape R.A.\ Jansen et al.\\[-6pt]
{\footnotesize Page \arabic{page}}\\[-14pt]}
\cfoot{\color{blue}\rule{\txw}{0.1pt}\\[-0pt]\hspace*{-2pt}\scriptsize\textsl{
Galaxy Assembly and SMBH/AGN-growth from Cosmic Dawn to the End of 
Reionization\hfill}}
\pagestyle{fancy}

\null

\noindent\hspace*{0.5in}\begin{minipage}[t]{6.0in}{\small 
{\bf Abstract:} \ We present a compelling case for a \emph{systematic} and
\emph{comprehensive} pan-chromatic (UV--near-IR) cosmological broad- and
medium-band imaging and grism survey that covers a wide area on the sky in
multiple epochs.  Specifically we advocate a tiered survey that covers
\tsim10\,deg$^2$ in two epochs to $m_{AB}$\,=\,28\,mag, \tsim3\,deg$^2$ in
seven epochs to $m_{AB}$\,=\,29\,mag, and \tsim1\,deg$^2$ in 20 epochs to
$m_{AB}$\,=\,30\,mag, each at 10$\sigma$ point source sensitivity.  Such a
survey will provide spectrophotometric redshifts accurate to
$\sigma_z/(1+z)\,\lesssim\,0.02$, faint source variability for
$\gtrsim$5$\times$10$^6$ galaxies and QSOs.  This survey is an essential
complement to \JWST\ surveys ($\lesssim$0.1\,deg$^2$ to
$m_{AB}\,\lesssim$\,31\,mag at $\lambda\!>$1100\,nm and $z\,\gtrsim\,$8). 
We aim to: 
(\textsl{1}) understand in detail how galaxies formed from the perturbations
in the primordial cosmological density field by studying faint \Lya-emitting
and Lyman-break galaxies at 5.5$\lesssim\,z\lesssim\,$8 and trace the
metal-enrichment of the intergalactic medium (IGM);
(\textsl{2}) measure the evolution of the faint end of the galaxy luminosity
function (LF) from $z$$\sim$8 to $z$$\sim$0 by mapping the ramp-up of
\ion{Pop}{2} star formation, (dwarf) galaxy formation and assembly, and
hence, the objects that likely completed the Hydrogen reionization at
$z\simeq\,$6;
(\textsl{3}) directly study the $\lambda\!<\,$91.2\,nm escape fractions of
galaxies and weak AGN from $z\sim\,$4.0--2.5, when the Helium reionization
in the universe finished;
(\textsl{4}) measure the mass- and environment-dependent galaxy assembly
process from $z\simeq\,$5 to $z\simeq\,$0, combining accurate
spectrophotometric redshifts with spatially resolved stellar populations and
kpc-scale structure for $\gtrsim$5$\times$10$^6$ galaxies;
(\textsl{5}) trace the strongly epoch-dependent galaxy merger rate and
constrain how Dark Energy affected galaxy assembly and the growth of
super-massive black holes (SMBHs);
(\textsl{6}) study $\gtrsim$10$^5$ weak AGN, including faint variable objects
that are likely feeding SMBHs in the faint-end of the QSO LF, over 10\,deg$^2$ 
and measure how the growth of SMBHs kept pace with galaxy assembly and spheroid
growth, and how this process was shaped by various feedback processes over
cosmic times since $z\sim\,$8.
The proposed study is not feasible with current instrumentation but argues
for a wide-field ($\gtrsim$250 arcmin$^2$), high-resolution
($\lesssim$0\farcs02--0\farcs11\,[300--1700\,nm]), UV--near-IR imaging
facility on a 4\,m-class space-based observatory.\\


} \end{minipage}

\bigskip

\noindent{\huge\bf O}ver the past decade, our knowledge about the universe  
at high redshifts has gradually extended to $z\!\simeq\,$6 with over a dozen
quasars discovered in the \SDSS\ at $z\,\gtrsim\,$6 and similar numbers of
\Lya\ emitters.  Of particular note are the discoveries of the first
``complete'' Gunn-Peterson troughs in the spectra of $z\!\!>$6 quasars and
the WMAP year-5 polarization measurement, which gives a 2$\sigma$ upper 
limit to the redshift range of the \ion{Pop}{3} star reionizing population
of $z\!\simeq\,$8--14. The reionization of the universe likely has left its
signature on the history of galaxy formation and evolution.  It is 
predicted to cause a drop in the cosmic star formation rate (SFR), and is
therefore accompanied by a dramatic fall in the number counts of objects
at $z\!\ge\,$6.

\smallskip

\noindent Since the UV shortward of $\lambda_0$=121.6\,nm is strongly
absorbed by intervening \ion{H}{1}, high redshift objects can be selected
using the so-called \textsl{drop-out} technique.  This technique requires
filters that bracket \Lya\ in the relevant redshift range.  Recent
\textsl{i}-band drop-out studies with \HST\ found significant numbers of
$z\!\simeq\,$6 candidates, although with non-negligible contamination by low
redshift elliptical galaxies and Galactic L- and T-dwarf stars.  As evidence
mounts that the Hydrogen reionization was largely complete by
$z\!\simeq\,$6, studies of the $z$\,=\,6--8 interval --- ``Cosmic Dawn''
---, will be of great cosmological importance.

\smallskip
\noindent{\color{blue}\rule{\txw}{0.1pt}\\[-4pt]\scriptsize\textsl{
Galaxy Assembly and SMBH/AGN-growth from Cosmic Dawn to the End of 
Reionization}}
\clearpage

%
\color{blue}
\noindent\leavevmode
\framebox[\textwidth]{
   \begin{minipage}[t][0.460\textheight][t]{\textwidth}
   \centering
   \includegraphics[width=0.425\textwidth]{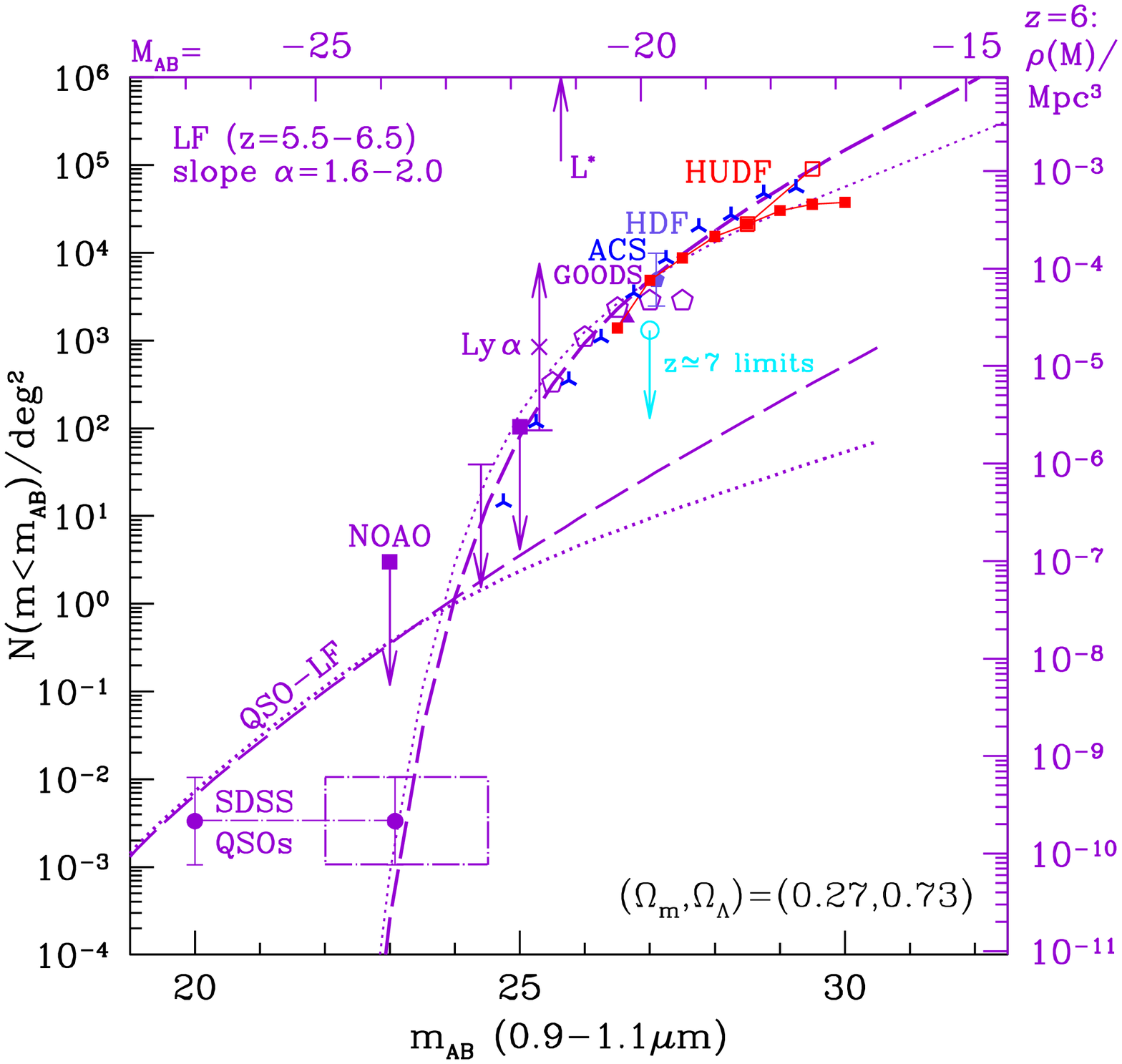} \ \ \ \ 
   \includegraphics[width=0.425\textwidth]{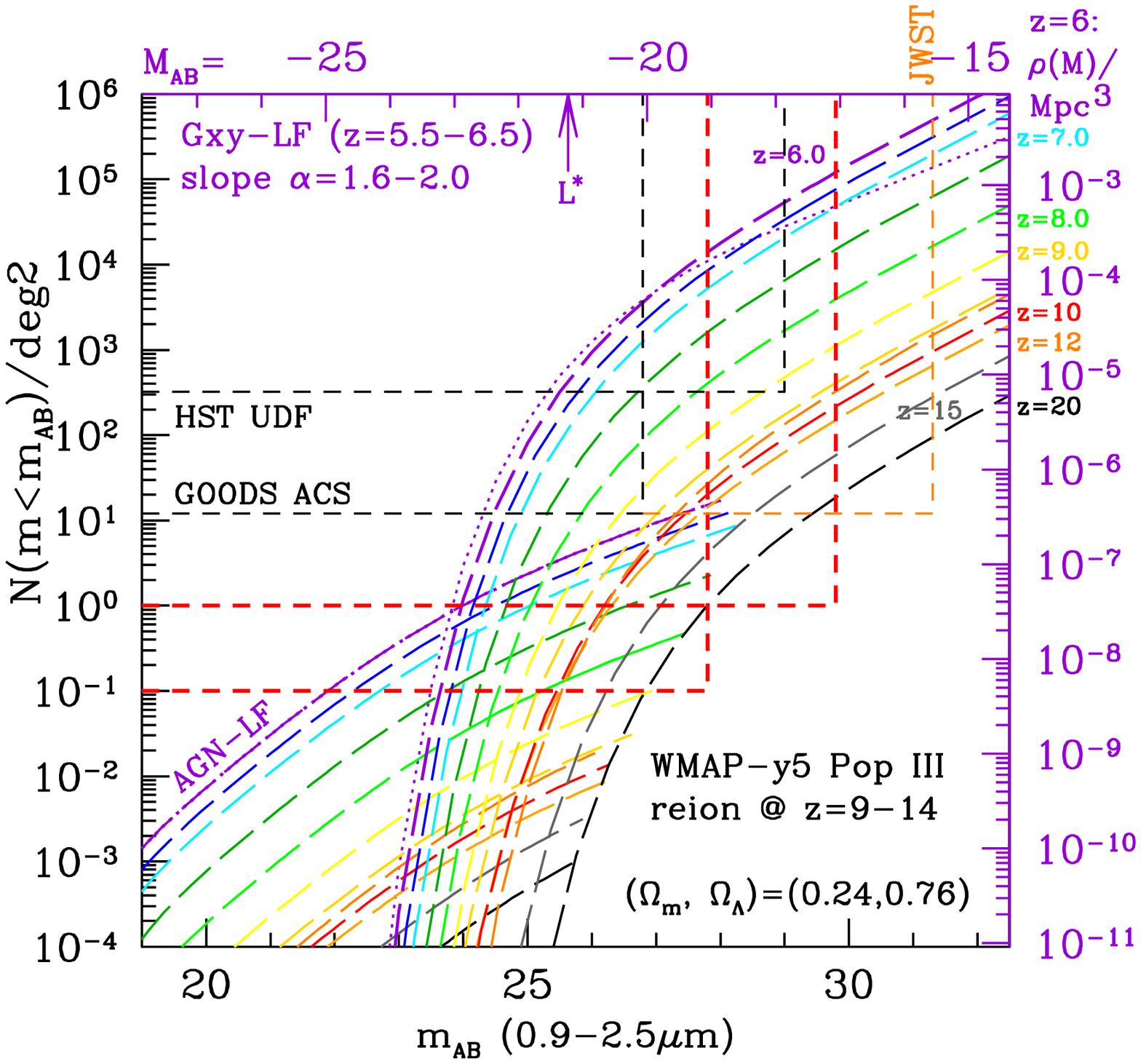}
   \parbox[t]{0.985\textwidth}{\vspace*{-12pt}\color{black}\small Fig.~1 ---
(\emph{a}) The integral luminosity function at $z\!\simeq\,$6 from various
samples.  Left and bottom axes give the observed surface densities and
fluxes, top and right axes absolute magnitudes and space densities.  The
surface density of $z\!\simeq\,$6 QSOs to AB=20\,mag from \SDSS\ and its
extrapolation to fainter fluxes using the measured faint-end slope of the
QSO LF ($|\alpha|\!\simeq\,$1.6) are also shown.  The faint-end slope of 
the galaxy LF is significantly steeper: $|\alpha|\!\simeq\,$1.8--2.0 
(long-dashed curve).  
(\emph{b}) The predicted surface density of $z\!\simeq\,$7--8 objects,
based on the very few candidates from the \HST/NICMOS HUDF surveys
(\emph{light blue upper limit in panel \emph{a}}) and constrained by the
total optical depth $\tau$ from WMAP, may be an order of magnitude lower
than that at $z\!\simeq\,$6, necessitating the very large survey areas
proposed here (\emph{red dashed limits}).}
   \end{minipage}
}
\color{black}
%

\smallskip

\noindent For how galaxies formed from perturbations in the primordial
density field, reflected in the Cosmic Microwave Background (CMB), remains a
major problem.  While numerical simulations can predict the formation of
dark matter halos and their clustering, the formation of stars that render
these halos visible is a complex process and hard to predict \textsl{a
priori}.  Thus, there is a great need to study galaxies observationally, at
all redshifts.  This is especially true at $z\,\gtrsim\,$6, where two major
changes took place: (\textsl{1}) metal enrichment of the intergalactic
medium (IGM), which must have occurred at $z\,\gtrsim\,$6 given the
observations of IGM metals even at $z$=5.7, and (\textsl{2}) reionization of
hydrogen in the IGM.  Since metallicity and ionization of gas changes the
nature of star formation by changing the available cooling mechanisms, it is
\emph{crucial} to push back our discovery of galaxies to $z\!>\,$6. 

\smallskip

\noindent Surveys for galaxies at $z\,\gtrsim\,$7 are very difficult for
many reasons, however.  The galaxies are fainter, both because of
cosmological dimming and also because of smaller characteristic luminosities
and sizes, resulting in low object surface densities (e.g., Fig.\,1).  It is
also important to realize that high redshift galaxy formation is
\emph{biased}, resulting in strong spatial variations in number density. 
For these reasons one would need to survey a large area (at least several
deg$^2$).  These searches need to be performed at
$\lambda\,\gtrsim\,$975\,nm, near and beyond the cut-off of Si CCDs.  In the
near-IR, there is a tremendous advantage of going to space, with its
$>$100--1000 times darker sky background. 

\smallskip

\noindent One class of primordial galaxies is easily identified in narrow-
or medium-band surveys from their strong, narrow \Lya\ emission and their
diminished flux blueward of this emission.  Indeed, \Lya-emitter surveys
have proven to be \emph{the} most successful technique to find galaxies at
the earliest cosmic epochs.  While the Gunn-Peterson troughs are produced by
neutral fractions of only 10$^{-4}$ or 10$^{-2}$ (for a homogenous or a
clumpy IGM, respectively), the change in number density of \Lya-emitters as
a function of redshift traces neutral fractions of the IGM of
$\gtrsim$30--80\%.  A quantitative study based on this principle requires
statistical samples of \Lya\ galaxies in each redshift bin.\linebreak
\noindent Ground-based surveys are and will remain severely limited in the
volume they can sample due to the necessity to use very narrow bandpass
filters (\tsim0.1\%) to observe between the strong atmospheric OH lines,
which makes them vulnerable to cosmic variance. 

\bigskip

%
\begin{wrapfigure}[22]{r}{0.50\textwidth}
   \color{blue}\vspace*{-15pt}\framebox[0.500\textwidth]{
      \begin{minipage}[t][0.470\textheight][t]{0.485\textwidth}
         \centering
         \includegraphics[width=0.85\textwidth]{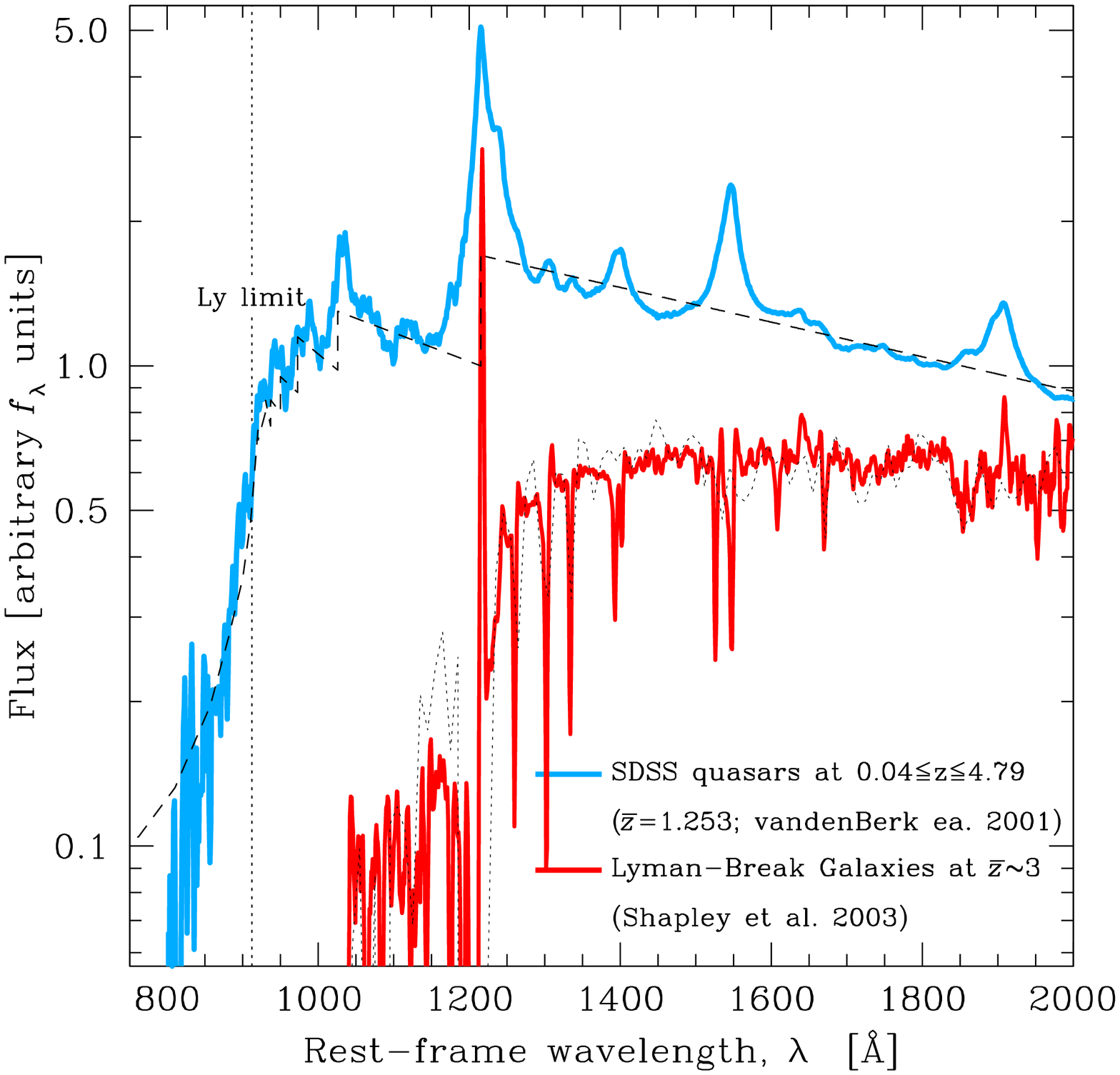}  
         \parbox[t]{\textwidth}{\vspace*{-8pt}\color{black}\small
Fig.~2 --- Whether the numerous dwarf galaxies or the much rarer AGN
finished reionization at $z\!\simeq\,$6 depends critically on the amplitude
and faint-end slope of the LF for each population.  The steep LF of dwarf
galaxies at $z\sim6$ could provide enough ionizing photons to complete
reionization if $f_{esc}\,\gtrsim\,$10\%.  Observations in UV--blue
filters in the proposed survey will yield escape fractions for
$\gtrsim$5$\times$10$^6$ objects. 
      }
      \end{minipage}
   }\color{black}
\end{wrapfigure}
%

\noindent The Hubble Ultra-deep Field (HUDF; Beckwith \etal 2006), our
deepest view of the distant universe yet, was collected over 4 epochs that
were each \tsim1 month apart.  Since the data of each seperate epoch still
reaches to \tsim28.0\,mag, this offered the unique opportunity to study the
variability of faint objects on time scales of months, corresponding to 4--5
weeks in the rest-frame.  Variability on such time scales betrays the
presence of a feeding SMBH within a galaxies' active nucleus (AGN).  The
redshift distribution of ``tadpole'' galaxies (galaxies in a particular
early-merger stage; the cover page shows examples) and variable objects in
the HUDF appear to be similar and likely holds clues to the mystery of how
the growth of spheriods and SMBHs has kept pace with the process of galaxy
assembly and resulted in the rather tight Magorrian-relation observed in the
local universe.  The present statistics are inadequate, however, and the
available redshift estimates imprecise.  A deep, multi-epoch survey over
$\gtrsim$1\,deg$^2$ would allow studying variability of faint objects over a
1000$\times$ larger area on the sky to similar depth, providing vastly
superior statistics. 

\bigskip

\noindent While a quantum leap forward, the past 15 years of \HST\ have
shown its results on the high-$z$ universe to still be severely limited by
its small field of view (FoV), limited aperture, and limited wavelength
range over which it provides high throughput. 
  %
  %
\noindent For a comprehensive study of the galaxy populations from the
height of the reionization epoch to the epoch were the present-day Hubble
sequence was established, one would require a space-based imaging facility
that provides:
\noindent\vspace*{-8pt}\begin{itemize}
\setlength{\itemsep}{-4pt}
\setlength{\itemindent}{-5.5ex}
\setlength{\labelwidth}{-5.5ex}

\item[\textsl{\bfseries(1)}] efficient wide-field coverage
($\gtrsim$250\,arcmin$^2$), sufficient to efficiently map areas large enough
to average out cosmic variance and find $z\,\gtrsim\,7$ objects with surface
densities $\gtrsim\,$0.1/deg$^2$;

\item[\textsl{\bfseries(2)}] high angular resolution, sufficient to
spatially resolve $\sim$1\,kpc sized objects at
0.5$\,\lesssim\,z\,\lesssim\,$8 at restframe wavelengths
$\lambda_0\!>$121.6\,nm (the lower-right figure on cover page demonstrates
that many objects fainter than $m_{AB}\sim27$ are no longer resolved by
\HST);

\item[\textsl{\bfseries(3)}] sufficient sensitivity to sample both the
bright and faint ends of the galaxy, the QSO, \Lya-emitter and Lyman-break
luminosity functions from $z\!\simeq\,$8 to $z\!\simeq\,$1, and to
$z\!\simeq\,$0 for the Balmer or 400\,nm breaks; \ \textsl{and}

\item[\textsl{\bfseries(4)}] a sufficiently rich complement of
near-UV--near-IR broad- and medium-band filters to provide photometric
redshift estimates accurate to $\sigma_z/(1+z)\,\lesssim\,0.02$ and to allow
efficient detection of \Lya-emitters from $z\!\simeq\,$8 to
$z\!\simeq\,$5.5\,.
\end{itemize}

\clearpage

\noindent We propose a near-UV--near-IR cosmological broad- and medium-band
imaging and grism survey that covers a wide area on the sky in muliple
epochs.  Specifically we advocate a tiered survey of \tsim10\,deg$^2$ in two
epochs to $m_{AB}$\,=\,28\,mag, \tsim3\,deg$^2$ in seven epochs to 
$m_{AB}$\,=\,29\,mag, and \tsim1\,deg$^2$ in 20 epochs to
$m_{AB}$\,=\,30\,mag, each at 10$\sigma$ point source sensitivity.  The use
of complementary deep, medium-deep, and wide surveys is a proven strategy to
maximize the scientific return for the investment in telescope time.  Such
surveys would provide spectrophotometric redshifts accurate to
$\sigma_z/(1+z)\,\lesssim\,0.02$, faint source variability for
$\gtrsim$5$\times$10$^6$ galaxies and QSOs, and a probe of the universe at
Cosmic Dawn when less than half of the hydrogen had been ionized.  It would
constitute an essential complement to deeper \JWST\ surveys
$m_{AB}\,\lesssim$\,31\,mag at $\lambda\!>$1100\,nm and $z\,\gtrsim\,$8)
over far smaller areas ($\lesssim$0.1\,deg$^2$). 

\medskip

\noindent In the following, we summarize our goals for each of the themes
of this survey. 

\bigskip

\noindent {\bfseries Key scientific themes that have arisen from recent
advances}

\smallskip

\noindent {\bfseries\itshape Evolution of the Faint-end Slope of the Dwarf
Galaxy Luminosity Function} \ The faint-end slope of the galaxy LF is
systematically steepening at higher redshifts, reaching a slope
$|\alpha|$=1.8--2.0 at $z\sim$6.  This implies that dwarf galaxies
collectively could have produced a sufficient number of ionizing photons to
complete the reionization of Hydrogen in the universe by $z\sim$6.  This
critically depends on the escape fraction, $f_{esc}$, of far-UV photons from
faint dwarf galaxies.  The proposed survey, in particular the UV--blue
broad-band filters, could answer this question for statistically meaningfull
samples per redshift bin.  It furthermore depends on the evolution of the
amplitude of the dwarf galaxy LF and whether or not there could be a
significant scatter in the faint-end slope due to clustering.  The surface
density of $z\!>$7 objects may be an order of magntide lower than that at
$z\!\sim\,$6, but the proposed surveys cover a sufficiently wide area to
unambiguously answer these questions. 

\smallskip

\noindent {\bfseries\itshape Tracing the Reionization History using
\Lya-Emitters} \ Observations sofar have failed to settle the issue of
whether the amplitude of the \Lya-emitter LF changes between $z$=5.7 and
$z$=6.5, or as extrapolated from the \emph{single} detection of a
\Lya-emitter at $z$=6.96\,.  The proposed medium-band surveys will derive
their LF as a function of redshift at $z\,\gtrsim\,$5.5 over a wide area for
large statistical samples and definitively address how the reionization of
the IGM progressed over time.  Furthermore, the data will allow measuring
the ages and clustering properties of \Lya-emitters, and, via the faint-end
slope of their LF, their contribution to the budget of ionizing photons. 
The latter is a complementary probe of cosmic reionization compared to the
counting experiment (LF amplitude).  This science is not addressed well by
many of the \emph{JDEM} concepts currently circulating, which are restricted
to imaging only in \emph{broad} bands.

\smallskip

\noindent {\bfseries\itshape Light Profiles of Dwarf Galaxies Around
Reionization} \ The average radial surface brightness profile derived from
stacked, intrinsically similar, $z\!\simeq\,$6, $z\!\simeq\,$5, and
$z\!\simeq\,$4 objects extracted from the \HST/ACS HUDF show a deviation
from a Sers\'{\i}c profile at progressive larger radii.  If interpreted as a
virial radius, in a hierarchical growth scenario, this would imply dynamical
ages for these dwarf galaxies of a 0.1--0.2\,Gyr at $z\!\simeq\,$6--4. 
These `dynamical' limits to their ages are comparable to age estimates based
on their SEDs, suggesting that the starburst that \emph{finished} the H
reionization at $z\!\simeq\,$6 may have started by a global onset of
\ion{Pop}{2} star formation at $z\!\simeq\,$6.5--7, or $\lesssim\,$200\,Myr
before $z\!\simeq\,$6.  The proposed surveys will yield light profiles,
color gradients, and dynamical states of $\gtrsim\,$10$^5$ dwarf galaxies at
0.5$\,\lesssim\,z\,\lesssim\,$7, and provide constraints to their ages from
their SEDs and, for a subset, also from systematic profile deviations. 

\smallskip

\noindent {\bfseries\itshape Lyman-continuum Escape Fraction of Dwarf
Galaxies and Weak AGN} \ A $z\!\simeq\,$6, the Lyman-continuum escape
fraction is likely somewhat larger than the 10--15\% measured for
Lyman-break galaxies at $z\!\simeq\,$3--4, reflecting the lower metallicity
at larger redshifts.  If indeed dwarf galaxies, and not QSOs, dominated the
late stages of reionization, then these objects cannot have started shining
pervasively much before $z\!\simeq\,$7--8, or no neutral \ion{H}{1} would
have been detected in front of $z\!\simeq\,$6 \SDSS\ quasars.  Hence, one
would expect to find a down-turn in their LF amplitude at $z\,\gtrsim\,$6.5
--- or a rapid onset of the cosmic SFR from $z\!\simeq\,$8 to
$z\!\simeq\,$6, which may be identified with the onset of dwarf galaxy
formation.  The proposed surveys will provide a unique glimpse into this
era of `Cosmic Dawn', where the first global IMF of \ion{Pop}{2} stars in
dwarf galaxies started forming.

\smallskip

\noindent {\bfseries\itshape The Process of Hierarchical Galaxy Assembly} \
The process of galaxy assembly may be directly traced as a function of mass
and cosmic environment in the redshift range
0.5$\,\lesssim\,z\,\lesssim\,$5.  The \HST\ Deep Fields have outlined how
galaxies formed over cosmic time, by measuring the distribution over
structure and type as a function of redshift.  Sub-galactic units appear to
have rapidly merged from $z\!\simeq\,$6--8 to grow bigger units to
$z\!\simeq\,$1.  Galaxies of all types formed over a wide range of cosmic
time, but with a notable transition around $z\!\sim$1.0\,.  Merger products
started to settle as galaxies with familiar morphologies, and evolved mostly
passively since then.  The fine details of this process still elude the
\HST\ surveys, because of inadequate spatial sampling and/or depth, and
because its FoV is too small to provide sufficient statistics (with the
exception of the COSMOS survey, but those observations were through only a
single filter).  The proposed imaging through multiple near-UV--near-IR
filters and grism(s) would yield robust spectrophotometric redshift
estimates ($\sigma_z/(1+z)\,\lesssim\,$0.02) for $\gtrsim\,$5$\times$10$^6$
galaxies with $m_{AB}\,\lesssim\,$28--30\,mag, and allow an analysis of
their stellar populations (through population synthesis modeling) and their
structure on spatial scales $\lesssim\,$few kpc. 

\smallskip

\noindent {\bfseries\itshape The Epoch-dependent Merger Rate of Galaxies} \
With robust photometric redshift estimates, it has become feasible to
meaningfully trace the pair fraction and galaxy major merger rate to very
faint limits ($m_{AB}\,\gtrsim\,$27\,mag).  From \HST/ACS flux limits and
panchromatic SED fitting, the currently available surveys have shown a mass
completeness limit for $z\,\lesssim\,$2--4 for
M$\,\gtrsim\,$10$^{10.0}$\,M$_{\odot}$ for primary galaxies in a pair and
M$\,\gtrsim\,$10$^{9.4}$\,M$_{\odot}$ for secondary galaxies.  The proposed
surveys would allow mapping the \emph{entire} epoch-dependent merger history
to at least 3\,mag fainter.  This would yield the galaxy merger density as a
function of total mass, mass ratio, redshift, and local overdensity and do
so for $\gtrsim$10$^6$ galaxies at $m_{AB}\,\lesssim\,$28--30\,mag over a
much wider range of masses
(10$^{0.8}$\,M$_{\odot}\,\lesssim\,$M$\,\lesssim\,$10$^{11.5}$\,M$_{\odot}$)
and for redshifts 0$\,\lesssim\,z\,\lesssim\,$7. 

\smallskip

\noindent {\bfseries\itshape The Growth of Super-Massive Black Holes} \
Through a multi-epoch variability study, the proposed surveys will be able
to measure the weak AGN fraction in $\gtrsim\,10^5$ field galaxies to
$m_{AB}\,\lesssim\,$28--30\,mag at $z\,\lesssim\,$8 directly, and so
robustly constrain how exactly growth of spheroids and SMBHs kept pace with
the process of galaxy assembly.  The panchromatic imagery and robust
spectrophotometric redshifts will allow decomposition of the AGN light from
that of the underlying galaxy.  This science theme also relies on a stable
PSF and proper PSF sampling.

\bigskip
\bigskip

\noindent{\bfseries Key Advances in Observation Needed}

\noindent \emph{Resolution} --- $\lesssim$0\farcs02--0\farcs11 
[300--1700\,nm] resolution is required in order to spatially resolve kpc-sized
objects at 0.5$\,\lesssim\,z\,\lesssim\,$8 at rest-frame wavelengths 
$\lambda_0\!>$121.6\,nm.

\noindent \emph{Wavelength agility} --- pan-chromatic wavelength coverage
from near-UV through near-IR for a comprehensive understanding of the
star-formation and assembly histories of galaxies, and to access \Lya\
emission redshifted to $z\sim\,$8.

\noindent \emph{Wide-field focal plane arrays} --- these are presently not
at sufficiently high TRL; investment is needed to improve yields, provide
cheaper devices and high-throughput assembly and testing to enable economies
of scale.  Such an investment would not just benefit the science proposed  
here.

\noindent \emph{Coatings} --- an investment in improving the relatively poor
broad-band performance of optical coatings of telescope mirrors in the UV,
with typical reflectances below 85\% (Al+MgF$_2$) directly results in a 
large increase in throughput for a given telescope aperture, or more
affordable missions for a given sensitivity requirement. 

\noindent \emph{Dichroics} --- most photons collected by telescopes are
rejected by bandpass filters.  Dichroic(s) potentially double (or even
triple) the observing efficiency of astronomical observatories (\eg,
\emph{Spitzer}/IRAC) and allow tuning downstream optics and detectors for
more optimal performance, avoiding compromises inherent in forcing
performance over more than an octave in frequency. 

\bigskip

\noindent{\bfseries Enabling science investigations}

\noindent The proposed science in the present white paper does not stand 
alone, but must build on a strong understanding of the physics of the 
star formation process in various environments, theoretical insights in
cosmological models of reionization and structure growth, as well as
synergy with both higher-resolution near-IR AO observations with  
next-generation giant-aperture telescopes, and deeper observations in
the near- and mid-IR with \JWST\ over small fields of view.  Also,
investment in human capital and in ground-based supporting and
path-finding programs, including operational support, should not be
ignored, as the overal science return of this and many `high-end'
programs critically depends on it.

\bigskip

\noindent{\bfseries Four central questions to be Addressed}

\vspace*{-8pt}\noindent\begin{itemize}
\setlength{\itemsep}{-4pt}
\setlength{\itemindent}{-5.5ex}
\setlength{\labelwidth}{-5.5ex}

\item[\textsl{\bfseries(1)}] How did reionization progress during the era of
`Cosmic Dawn'? Was it an extended, a rather abrupt, or even a multiple event?

\item[\textsl{\bfseries(2)}] How did the faint end of the galaxy luminosity
function evolve from the onset of \ion{Pop}{2} star formation till the end
of the reionization epoch?

\item[\textsl{\bfseries(3)}] How exactly did AGN and SMBH growth keep pace
with the process of galaxy assembly?  How did AGN growth decline with the
galaxy merger rate and the cosmic SFR?

\item[\textsl{\bfseries(4)}] Was there indeed an epoch of maximum merging and
AGN activity around $z\!\simeq\,$1--2 for the more massive galaxies, before
the effects from the increasingly dominant Dark Energy kicked in?  How does
this peak epoch depend on galaxy total mass or bulge mass, and (how) does 
this support the galaxy downsizing picture?

\end{itemize}

\bigskip

\noindent{\bfseries Area of Unusual Discovery Potential for the Next Decade}

\noindent Combination of a large collecting area, very wide field of view, 
high angular resolution, wavelength agility and/or multiplexing advantage  
would allow orders of magnitude more efficient UV--optical observations of
star formation, galaxy assembly, and SMBH-growth processes and, moreover,
open up a new domain in discovery space near and far.
  %
  %
Injection into L2 (or Earth Drift-Away) orbits allows provide dynamical and
thermal stability, and increases (doubling) in efficiency over LEO orbits
and, hence, lower cost per hour of observation (all other variables being
equal).
  %
  %
Large focal plane array (dozens to hundreds of individual CCD or CMOS
detectors) and dichroic camera (simultaneous observation in two or more
channels of the same field of view) technology is better matched to the
collimated beams provided by optical telescope assemblies and less
wasteful in terms of collected photons, maximizing science output and
especially benefitting survey science with a lasting legacy beyond the
nominal duration of a mission.
  %
  %
Survey science allows discovery of very rare objects amongs billions and
billions, the positions an properties of which may not be knowable a priori.

\vspace*{0.5in}

{\small
\noindent{\bfseries References}\\[-1pt]
Beckwith, S., Stiavelli, M., Koekemoer, A., \etal 2006, AJ 132, 1729
}

\end{document}